\documentstyle[fleqn, epsfig, a4]{article}  
\date{}
\newcommand{\AmS}{{\protect\the\textfont2  
   A\kern-.1667em\lower.5ex\hbox{M}\kern-.125emS}}  
   
\newcommand{\Dlr}{\stackrel{\leftrightarrow}{D}}  
\newcommand{\Dl}{\stackrel{\leftarrow}{D}}  
\newcommand{\Dr}{\stackrel{\rightarrow}{D}}

\newcommand{\pslash}{{\not{\hspace{-0.08cm}p}}}  
\newcommand{\sslash}{{\not{\hspace{-0.07cm}s}}}  
\newcommand{\half}{\frac{1}{2}}  
\newcommand{\calO}{ {\cal O} }

\begin{document}

\title{
\vspace{-1.0cm}
\flushleft{\normalsize DESY 99-115} \hfill\\
\vspace{-0.65cm}
\flushleft{\normalsize TPR 99-15} \hfill\\
\vspace{-0.65cm}
\flushleft{\normalsize HUB-EP-99/41} \hfill\\
\vspace{-0.65cm}
\flushleft{\normalsize HLRZ 99-36} \hfill\\
\vspace{-0.55cm}
\flushleft{\normalsize August 1999} \hfill\\ 
\vspace{0.5cm}
\centering{\LARGE \bf Operator improvement for Ginsparg-Wilson fermions}} 
\author{\Large S.~Capitani$^1$, M.~G\"ockeler$^2$, R.~Horsley$^3$,   
 P.E.L.~Rakow$^2$, G.~Schierholz$^{1,4}$\\[2.5ex]  
  $^1$ Deutsches Elektronen-Synchrotron DESY, 
 D-22603 Hamburg, Germany\\[0.75ex]
  $^2$ Institut f\"ur Theoretische Physik, Universit\"at Regensburg,\\  
 D-93040 Regensburg, Germany\\[0.75ex]  
  $^3$ Institut f\"ur Physik, Humboldt-Universit\"at zu Berlin,  
 D-10115 Berlin, Germany\\[0.75ex]  
  $^4$ Deutsches Elektronen-Synchrotron DESY,\\
    John von Neumann-Institut f\"ur Computing NIC,
  D-15735 Zeuthen, Germany}  
    
 \maketitle  
   
 \begin{abstract}  
 The improvement of fermionic operators for Ginsparg-Wilson fermions
 is investigated. We present explicit formulae for improved Green's functions,
 which apply both on-shell and off-shell.  
 \end{abstract}  
   

 \section{Introduction}  
   
  Recently, there has been a great deal of interest in
 Ginsparg-Wilson fermions~\cite{GW}, because it is now realised that they 
 allow the calculation of chirally symmetric physics without 
 having any doubling problem~\cite{Hasenfratz,Neuberger,Luescher}. We
 also know that because of its chiral properties the Ginsparg-Wilson fermion
 matrix is automatically an $O(a)$ improved action \cite{NiederReview},
 in exactly the same sense that the clover action is an improved action.

  If we are interested in going beyond spectrum calculations 
 to compute improved matrix elements, for example for structure
 functions and decay constants, we also need to know how to improve fermion
 operators. When we compute forward hadronic matrix elements, it is
 enough if the
 operators are improved for on-shell quantities. However, some
 methods of doing non-perturbative renormalisation \cite{Martinelli} require
 calculations of off-shell Green's functions, with a virtuality
 large enough that
 we can reasonably compare with continuum perturbation theory.
 To do this well, we would like to be able to remove $O(a)$ effects
 from off-shell Green's functions too. 

  In this paper we find several ways of improving operators 
 for the Ginsparg-Wilson action, 
 both for on-shell and off-shell Green's functions.  
 The paper is organised as follows. In sect.~2 we first study the
 fermion propagator, and in sect.~3 we test the results in the free
 case for a particular realisation of Ginsparg-Wilson fermions. In
 sect.~4 we then turn to our main subject, operator improvement. The
 major outcome is that, in the simplest form, the improved operators 
 do not require any coefficients to be tuned non-perturbatively.
 In sect.~5 we briefly comment on the Ward identities, and in sect.~6
 we conclude. 
  
 \section{Fermion propagator}

   In this section we review some known results concerning the 
 Ginsparg-Wilson propagator, as preparation for our consideration 
 of bilinear fermionic operators. 
 The basic Ginsparg-Wilson condition is \cite{GW} 
 \begin{equation}
 D_{GW} \, \gamma_5 + \gamma_5 \, D_{GW}    
  = a \, D_{GW} \, \gamma_5  \, D_{GW},
 \label{GW_condition} 
 \end{equation} 
 where $D_{GW}$ is the Ginsparg-Wilson fermion matrix. The fermion
 matrix does not anti-commute with $\gamma_5$, but there is
 nevertheless a form of chiral symmetry present \cite{Luescher}. 

\begin{figure}[tbh]
\begin{center}
\epsfig{file = 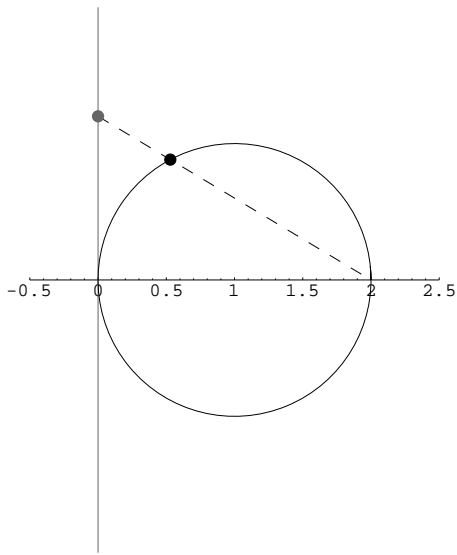, width = 6cm}
\end{center} 
\caption{ 
 The spectrum of the Ginsparg-Wilson matrix $D_{GW}$ 
 (the circle) and of the associated matrix $K_{GW}$
 (dotted vertical line).  
 The dashed line links an eigenvalue of $D_{GW}$ to the 
 corresponding eigenvalue of $K_{GW}$. }
\label{fig:gw}
\end{figure}

  Let us now look more closely at the Ginsparg-Wilson matrix. 
 From the  matrix $D_{GW}$ we can 
 define a related matrix \cite{Zenkin} 
 \begin{equation}
 K_{GW} \equiv \left( 1 - \frac{a}{2} D_{GW}\right)^{-1} D_{GW} . 
 \label{kdef} 
 \end{equation}
 We will discuss later what happens when $D_{GW}$ has an 
 eigenvalue equal to $2/a$.
 This implies 
 \begin{equation}
   D_{GW} =  \left( 1 + \frac{a}{2} K_{GW}\right)^{-1} K_{GW} . 
 \label{dsub} 
 \end{equation}  
 The eigenvalues of $D_{GW}$ lie on a circle of radius $1/a$ 
 and centre $1/a$, while the eigenvalues of $K_{GW}$ lie on
 the imaginary axis. The relationship between the eigenvalues of
 the two matrices is shown in Fig.~\ref{fig:gw}. 
 If we substitute eq.~(\ref{dsub}) into the condition 
 (\ref{GW_condition}) we find that 
 \begin{equation} 
 K_{GW} \, \gamma_5 + \gamma_5 \, K_{GW} =  0 . 
 \end{equation}  
 The chiral properties of $K_{GW}$ are even closer to those of the 
 continuum Dirac operator than $D_{GW}$. 
 The fermion propagator we really want to use is the 
 propagator calculated with $K_{GW}$. The propagator 
 calculated from $D_{GW}$ has contact terms of $O(a)$, 
 which mean that if we Fourier transform it we will find
 $O(a)$ lattice artifacts when we are off-shell. 
 A propagator calculated from  $K_{GW}$  contains no
 contact terms: it satisfies chirality even at zero distance.
 Therefore it will be improved off-shell too. 

 We want to write down improved Green's functions for the massive 
 as well as for the massless theory.  
  When we consider massive fermions, we will use the
 fermion matrix \cite{NiederReview} 
 \begin{equation}
 M \equiv \left(1 - \frac{1}{2} a m_0 \right) D_{GW} + m_0 \, .
 \end{equation} 
 This is the usual choice when working with Ginsparg-Wilson fermions, 
 and it has the advantage of giving a
 simple linear relationship between bare and renormalised 
 fermion masses. Most other ways of adding a quark mass lead to
 non-linear relationships of the form
 $m_R \propto m_0 (1 + a m_0 b_m + \cdots) $.
 However, despite the slightly more complicated form of the answer, 
 the improvement schemes we suggest in this paper
 also work with minor modifications 
 for other choices of $M$, for example $M = D_{GW} + m_0$.  

   The fermion propagator    
 \begin{equation}
 S \equiv \frac{1}{a^4} \left\langle M^{-1} \right\rangle 
 \end{equation}
 still contains contact terms violating chiral symmetry.%
 \footnote{ 
    We will be considering Green's functions involving charged 
  or coloured fermions. These are, of course, only defined if the 
  gauge is fixed in some way, so expectation values are always 
  to be understood as averages over gauge configurations with 
  some gauge-fixing term present. Which gauge is chosen makes
  no difference to any of the results in this paper, as all the 
  improvement coefficients are gauge independent.} 
 The fermion propagator with the correct chiral symmetry
 properties is the propagator corresponding to $K_{GW}$, i.e.
 \begin{equation}
 S_\star = \frac{1}{a^4} \left\langle \frac{1}{K_{GW} +m_0}\right\rangle . 
 \label{Sheur} 
 \end{equation} 
  We will always use $\star$ to denote improved quantities.
 The improved Green's functions 
 have the correct chiral properties, and we therefore know 
 that they are free of $O(a)$ discretisation errors. This is because
 any $O(a)$ term in a Green's function has the opposite
 chirality to the leading term, so checking chirality is a simple
 way of testing for $O(a)$ discretisation errors. 
 The improved propagator (\ref{Sheur}) can be calculated in terms of the 
 fermion matrix $M$. Inserting  the definition (\ref{kdef}) 
 into eq.~(\ref{Sheur}) we find 
 \begin{equation}
 S_\star (x,y) = \frac{1}{1 + a m_0 b_\psi }
 \left( S(x,y) - \frac{a}{2} \lambda_\psi \delta (x-y) \right), 
 \label{Sstardef} 
 \end{equation} 
 with improvement coefficients 
 \begin{eqnarray}
 b_\psi &=&  - \, \frac{1}{2}, \\
 \lambda_\psi &=& 1 . 
 \label{propimpco} 
 \end{eqnarray} 
    We use the lattice delta function
  \begin{equation}
  \delta(x-y) \equiv \frac{1}{a^4} \delta_{x y},
  \end{equation} 
  where $\delta_{x y}$ is the dimensionless Kronecker delta function. 
 Although the propagators $S$ and $S_\star$ in eq.~(\ref{Sstardef})
 are of course gauge-dependent, the improvement coefficients
 $b_\psi$ and $\lambda_\psi$ are gauge independent. 

 Note that although we have used $K_{GW}$ at intermediate stages in
 the discussion, our final result (\ref{Sstardef}) only involves
 inverting the fermion matrix $M$. As noted before, $K_{GW}$ is not
 well defined if $D_{GW}$ has eigenvalues exactly equal to $2/a$,
 which will happen in topologically non-trivial configurations. 
 On the other hand, the fermion matrix $M$ has no such problems.
 It is always invertible
 for $m_0 > 0$, so eq.~(\ref{Sstardef}) can be applied even in
 configurations with a non-trivial winding number. 
 Our improved Green's functions will always be of such a form 
 that only $M^{-1}$ appears as a propagator. 
 An identity equivalent to eq.~(\ref{Sstardef}) 
 that will often prove useful is 
 \begin{equation}
 \frac{1}{K_{GW}+m_0} = \left( 1- \frac{a}{2} D_{GW}\right) M^{-1}
 =  M^{-1}  \left( 1- \frac{a}{2} D_{GW}\right) . 
 \label{KDident} 
 \end{equation} 
 
  Now that we have defined a propagator, we can look for  
 an expression for the chiral condensate.  
   The natural choice of order parameter for chiral symmetry 
 is to take the trace of the improved propagator $S_\star$, 
 \begin{equation}
 \left\langle \bar{\psi}(x) \psi(x) \right\rangle_\star 
 = {\rm Tr} \, S_\star(x,x) 
 = \frac{1}{a^4} {\rm Tr} \left\langle \left(1 - \frac{a}{2} D_{GW} \right)
 M^{-1} \right\rangle \, . 
 \end{equation} 
 The final form of the improved chiral condensate is the same as
 that given in \cite{Chandra}. 

 \section{The fermion propagator in the free theory}

 In this section we consider one explicit realisation of the 
 Ginsparg-Wilson condition (\ref{GW_condition}), namely Neuberger's
 Dirac operator \cite{Neuberger}. 
 We show that the formulae we derived do indeed lead to 
 results free of $O(a)$ effects. 
 
  Starting from the massless Wilson fermion matrix $D_W$, Neuberger
 introduces the matrix $A$, defined by
 \begin{equation}
 A \equiv 1 - a D_W . 
 \end{equation} 
 It can then be shown that the operator 
 \begin{equation}
 D_N \equiv \frac{1}{a} \left( 1 - A / \sqrt{ A^\dagger A} \, \right) 
 \end{equation} 
 satisfies the Ginsparg-Wilson condition (\ref{GW_condition}) . 

 In the free theory, the Wilson matrix is diagonal in momentum, 
 and has the value
 \begin{equation}
 D_W(p) = {\rm i} \sslash +  W ,
 \end{equation} 
 where we use 
 \begin{eqnarray}
 W &=& \sum_\mu (1 - \cos a p_\mu ), \\
 \sslash  &=& \sum_\mu \gamma_\mu \sin a p_\mu, \\
 s^2 &=&  \sum_\mu \sin^2  a p_\mu . 
 \end{eqnarray} 
 Calculating $D_N$, the Ginsparg-Wilson matrix 
 corresponding to $D_W$, we find \cite{Luescher} 
 \begin{equation}
 A^\dagger A = (1 - W)^2 + s^2 
 \end{equation} 
 and
 \begin{equation} 
 D_N(p) = \frac{1}{a} \,\frac{{\rm i} \sslash 
 + \left( W - 1 + \sqrt{ (1-W)^2 + s^2}\right)}
 {\sqrt{ (1-W)^2 + s^2 }} . 
 \label{DN_val} 
 \end{equation} 
 If we expand $D_N(p)$ for small $p$, we find 
 \begin{equation}
 D_N(p) = {\rm i} \pslash + \half a p^2 + O(a^2 p^3) , 
 \end{equation} 
 so $D_N$ still has discretisation errors of $O(a)$. 

 However, when we calculate $K_N(p)$ according to the 
 formula (\ref{kdef}), we find 
 \begin{equation}
 K_N(p) = \frac{2}{a} \, \frac{ {\rm i} \sslash } 
 {1 - W +  \sqrt{ (1-W)^2 + s^2}} . 
 \end{equation} 
  Finally in $K_N(p)$ we have an operator which anti-commutes 
 with $\gamma_5$. When we expand it for small $p$, we find 
 \begin{equation}
 K_N(p) = \frac{ {\rm i} \sum_\mu \gamma_\mu 
 ( p_\mu - \frac{1}{6} a^2 p_\mu^3 + O(a^4 p^5) ) }
 { 1 - \frac{1}{4} a^2 p^2 + O(a^4 p^4) } 
 \end{equation} 
  which has lattice errors of $O(a^2)$. 

 However, $K_N$ does have problems of its own. It diverges  when 
 $s^2 = 0$ and $W \ge 1$, which occurs at the `doubler'
 momenta $p a = (\pi, 0,0,0), \cdots$.  Because it has poles
 in momentum space, it would also be
 an extremely non-local matrix if Fourier transformed back
 into position space. This is, of course, inevitable: 
 the Nielsen-Ninomiya theorem tells us that a fermion matrix
 that anti-commutes with $\gamma_5$ and has no doubling 
 must be non-local. 
 
\begin{figure}[tb]
\begin{center}
\epsfig{file = 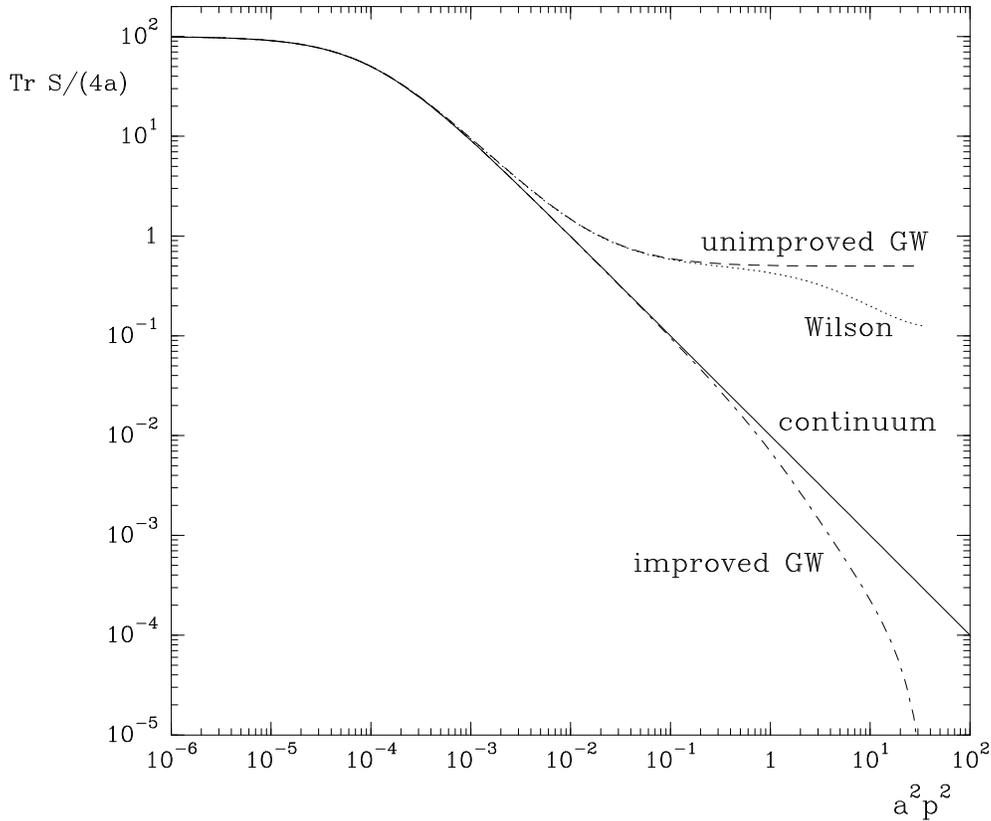, angle = 270, width = 13cm} 
\end{center} 
\vspace*{0.5cm} 
\caption{ 
 The scalar part of the free propagator, $(1/4 a) {\rm Tr} S(p)$,
 plotted against $a^2 p^2$, 
 for the various fermion matrices considered in this section. 
 The Wilson propagator is the dotted line, the unimproved propagator
 from $D_N$ is the dashed line, 
  the improved propagator $S_\star$ is the dot-dashed line, and 
 the solid line is the continuum result $m_0/(p^2 + m_0^2)$. 
 All curves are plotted for $a m_0 = 0.01$. 
 The momentum is taken in the direction $(1,1,1,1)$. 
  }
\label{fig:prop}
\end{figure}

 The improved propagator does not have these difficulties. It is 
 \begin{eqnarray}
 S_\star(p,m_0) &=& \frac{ -{\rm i} \sslash /a 
 + m_0 \left( 1 - W +  \sqrt{ (1-W)^2 + s^2} \right) /2} 
 {2 \left( W - 1 +  \sqrt{ (1-W)^2 + s^2} \right) /a^2 
 + m_0^2 \left( 1 - W +  \sqrt{ (1-W)^2 + s^2} \right) /2 } 
 \nonumber \\ &&  \nonumber \\
 &=& \frac{ -{\rm i} \pslash + m_0 }{ p^2 + m_0^2 } + O(a^2). 
 \label{Sstarfree} 
 \end{eqnarray} 
 The denominator has no zeroes if $m_0^2 > 0$, and if $m_0 =0$ the
 only zero is at $p = 0$. The numerator vanishes at the `doubler'
 momenta. If we expand this propagator as a power series in 
 $a$, we see that there is no $O(a)$ term: the discretisation errors
 of $S_\star$ are of $O(a^2)$. 

 In Fig.~\ref{fig:prop} we compare the trace of the various lattice
 propagators with the correct continuum result. 
 The Wilson propagator and the unimproved Ginsparg-Wilson
 propagator $M^{-1}$ 
 both deviate from the correct result when $a^2 p^2 \approx a m_0$, 
 but the improved propagator $S_\star$ is good up to $a^2 p^2 \approx 1$. 

 \section{Improvement of flavour non-singlet fermion operators
 \label{sect:impop} }

 Next we want to improve the Green's function corresponding to
 a flavour non-singlet operator 
\begin{equation}
{\cal O} = \bar{\psi} O \psi, 
\end{equation}
where $O$ can include Dirac structure and covariant derivatives. 
 Covariant derivatives can be represented by the usual formulae
 \begin{eqnarray}
  \Dr_\mu f &=& \frac{1}{2 a} \left[ U_\mu(x) f(x + \hat{\mu})
        - U_\mu^\dagger(x - \hat{\mu}) f(x - \hat{\mu}) \right], 
 \nonumber \\
 {\bar f} \Dl_\mu  &=& \frac{1}{2 a} \left[ 
  {\bar f}(x + \hat{\mu})  U_\mu^\dagger(x) 
 - {\bar f}(x - \hat{\mu})  U_\mu (x - \hat{\mu})
\right],
 \end{eqnarray} 
 or by any other expression with discretisation errors of $O(a^2)$. 
 We want our improved Green's function to be
 \begin{equation}
 G_\star^\calO = \frac{1}{a^4} \left\langle 
 \frac{1}{K_{GW} + m_0 } O  \frac{1}{K_{GW} + m_0 }
 \right\rangle \, . 
 \label{Ggoal} 
 \end{equation} 
 This will have discretisation errors of $O(a^2)$, 
 because the propagators used have been improved, and the 
 usual discretisation of $O$ is valid up to terms of $O(a^2)$.  

   We need  a formula analogous to eq.~(\ref{Sstardef}), 
 giving $G_\star^\calO$ in terms of $M^{-1}$. The simplest 
 way to construct such an identity is by using eq.~(\ref{KDident}), 
 which gives
 \begin{equation}
  G_\star^\calO = \frac{1}{a^4} 
  \frac{1} { 1 + a m_0 b_\psi }
 \left\langle M^{-1} \widetilde{O}  M^{-1} \right\rangle, 
 \label{Goffshell} 
 \end{equation}  
 where 
 \begin{eqnarray} 
  \widetilde{O} &=& ( 1 + a m_0 b_\psi ) 
 \left(1 - \frac{a}{2} D_{GW} \right) O \left(1 - \frac{a}{2} D_{GW} \right)
 \nonumber \\
  &=& O +  a m_0 b_\psi O 
 -\frac{a}{2}  ( 1 + a m_0 b_\psi )  ( D_{GW} O + O \, D_{GW} )
  + \frac{a^2}{4}  ( 1 + a m_0 b_\psi )  D_{GW} O\, D_{GW} . 
 \label{Otild} 
 \end{eqnarray} 
 Equation~(\ref{Goffshell}) is written with the same wave function 
 improvement factor as eq.~(\ref{Sstardef}), because the quantity
 of physical interest is always the ratio $G_\star / S_\star$, 
 and we want the wave function factors to cancel.

 Equation~(\ref{Goffshell}) is not the most general expression for 
 $G_\star^\calO$. We can use the equation of motion 
 \begin{equation}
 D_{GW} M^{-1} = - \, \frac{m_0}{1 - a m_0/2}  M^{-1}
 + \frac{1}{1 - a m_0/2} \delta_{x y} 
 \end{equation} 
 to show that this is equivalent to the expression
 \begin{equation}
  G_\star^\calO =  
  \frac{1} { 1 + a m_0 b_\psi }  
 \left[ G_\circ - \frac{a}{2} \lambda_\calO C^\calO
 + \frac{a^2}{4}  \eta_\calO \langle O \rangle \right], 
 \label{GW3pt_2} 
 \end{equation}
 where
 \begin{eqnarray} 
  G_\circ &\equiv & \left\langle 
 M^{-1} O_\star M^{-1} \right\rangle, \\
 O_\star &\equiv & O + a m_0 c_0 O 
 - \frac{a}{2} c_1 ( D_{GW} O + O \, D_{GW} ) 
 + \frac{a^2}{4} c_2  D_{GW} O\, D_{GW}, 
 \label{GWimpop2} \\ 
 C^\calO &\equiv & 
 \left\langle O M^{-1}\right\rangle + 
 \left\langle M^{-1} O \right\rangle ,
 \end{eqnarray}  
 with 
 \begin{eqnarray}
 c_0 &=& \frac{ \frac{1}{2} - c_1 }{1 - \frac{1}{2}a m_0 } 
 \, - \,\frac{ a m_0 c_2 } {4 (1-\frac{1}{2}a m_0)^2 }, \\ 
 \lambda_\calO &=& \frac{1 - c_1}{1 - \frac{1}{2}a m_0 } 
 \,- \,\frac{ a m_0 c_2 } {2 (1-\frac{1}{2}a m_0)^2 }, \\ 
 \eta_\calO &=& \frac{1 -c_2 - \frac{1}{2}a m_0}{(1 - \frac{1}{2}a
   m_0)^2 } \, .   
 \end{eqnarray} 
  There are two free parameters in this 
 system of equations. The improvement coefficients $c_1$ 
 and $c_2$ can take 
 any value, but once it is chosen, the values of the other
 improvement coefficients are fixed. This freedom comes from 
 the equations of motion, which allow us to compensate for a change
 in one of the improvement coefficients by adjusting the other
  coefficients. For an example of this in the clover action 
 see~\cite{Lat97}.        

   The terms proportional to $C^\calO$ and $\langle O \rangle$ 
 can be interpreted as contact terms. The Green's function we are
 interested in has the form
 $\langle \psi_i \bar{\psi_j} O_{j k} \psi_k \bar{\psi_l} \rangle$. 
 On the lattice we should expect to see a contact term of the 
 form $\delta_{i j} \langle O_{j k} \psi_k \bar{\psi_l} \rangle 
 + \langle \psi_i \bar{\psi_j} O_{j k} \rangle \delta_{k l}$, 
 with a coefficient of $O(a)$, and a `double contact term'
 of the form $\delta_{i j} \langle O_{j k} \rangle \delta_{k l}$,
 with a coefficient of $O(a^2)$.  

  If we are looking at the operator Green's function on-shell, 
 these contact terms are irrelevant, because the fermion fields 
 will all be well separated in position, and so the delta functions
 are all zero. In this case we can use any values of $c_1$ and $c_2$,
 as long as we use the correct $c_0$ value.  

   On the other hand, if we look at the off-shell Green's functions,
 we have to take the contact terms into account. One possibility
 would be that as well as computing $G_\circ$ 
 one would also compute $C^\calO$ and $\langle O \rangle$  
 and add them with the coefficients $\lambda_O$ and $\eta_O$. 
 A more elegant procedure would be to use
 eq.~(\ref{Goffshell}), i.e. to choose $c_1$ and $c_2$ 
 so that the contact terms are absent, and $G_\circ$ is 
 improved both on- and off-shell. The improvement coefficients
 for this special case are 
 \begin{eqnarray} 
 c_1 = c_2 &=& 1 - \frac{1}{2} a m_0, \nonumber \\
 c_0 &=& - \, \frac{1}{2} . 
 \end{eqnarray}  
   Note that, as in the previous section, the matrix $K_{GW}$ is
 only used heuristically. To find the improved Green's functions
 $S_\star$ and $G_\star^\calO$, the only matrix that we really
 have to invert is $M$. 

  Finally, we mention another way to calculate the improved Green's 
 functions $S_\star$ and $G_\star^\calO$ of
 eqs.~(\ref{Sstardef}), (\ref{Ggoal}).  From eq.~(\ref{KDident}) 
 we can write the improved propagator as 
 \begin{equation}
 S_\star = \left\langle \left( 1- \frac{a}{2} D_{GW}\right)
  M^{-1} \right\rangle . 
 \end{equation} 
 Since $D_{GW}$ and $M$ commute, this can be rewritten in the
 more symmetric form 
 \begin{equation} 
 S_\star =  \left\langle 
 \left( 1 - \frac{a}{2} D_{GW} \right)^{\frac{1}{2}} 
 M^{-1}  \left( 1 - \frac{a}{2} D_{GW} \right)^{\frac{1}{2}} \right\rangle.
 \label{Srot} 
 \end{equation} 
 This formula gives exactly the same improved propagator as 
 eq.~(\ref{Sstardef}), but it lends itself to a somewhat different
 interpretation. In eq.~(\ref{Sstardef}) we subtract an unwanted 
 contact term present in the unimproved propagator, while 
 eq.~(\ref{Srot}) can be interpreted as the propagator of a 
 `rotated' fermion field, which is similar to the picture of 
 clover improvement presented in  \cite{Heatlie}. 

 Similar formulae can be written for the improved Green's
 functions, where instead of adding irrelevant higher dimension terms
 to $O$  we perform a `rotation' of the fermion fields. 
 The resulting formula is 
 \begin{equation}
 G_\star^\calO = \left\langle
  \left( 1 - \frac{a}{2} D_{GW} \right)^{\frac{1}{2}} M^{-1} O^{\rm rot} 
 M^{-1}  \left( 1 - \frac{a}{2} D_{GW} \right)^{\frac{1}{2}}  \right\rangle,
 \label{Grot} 
 \end{equation} 
 where 
 \begin{equation}
  O^{\rm rot} \equiv  \left( 1 - \frac{a}{2} D_{GW} \right)^{\frac{1}{2}}
 O  \left( 1 - \frac{a}{2} D_{GW} \right)^{\frac{1}{2}} . 
 \end{equation}  
 Using the identity (\ref{KDident}) it is easy to see that 
 eq.~(\ref{Grot}) is equivalent to eq.~(\ref{Ggoal}). 
 The rotation approach requires us to take the square root of a matrix, 
 which is not needed in the irrelevant operator approach, so it might 
 be more costly to implement. 
 
\begin{figure}[tb]
\begin{center}
\epsfig{file = 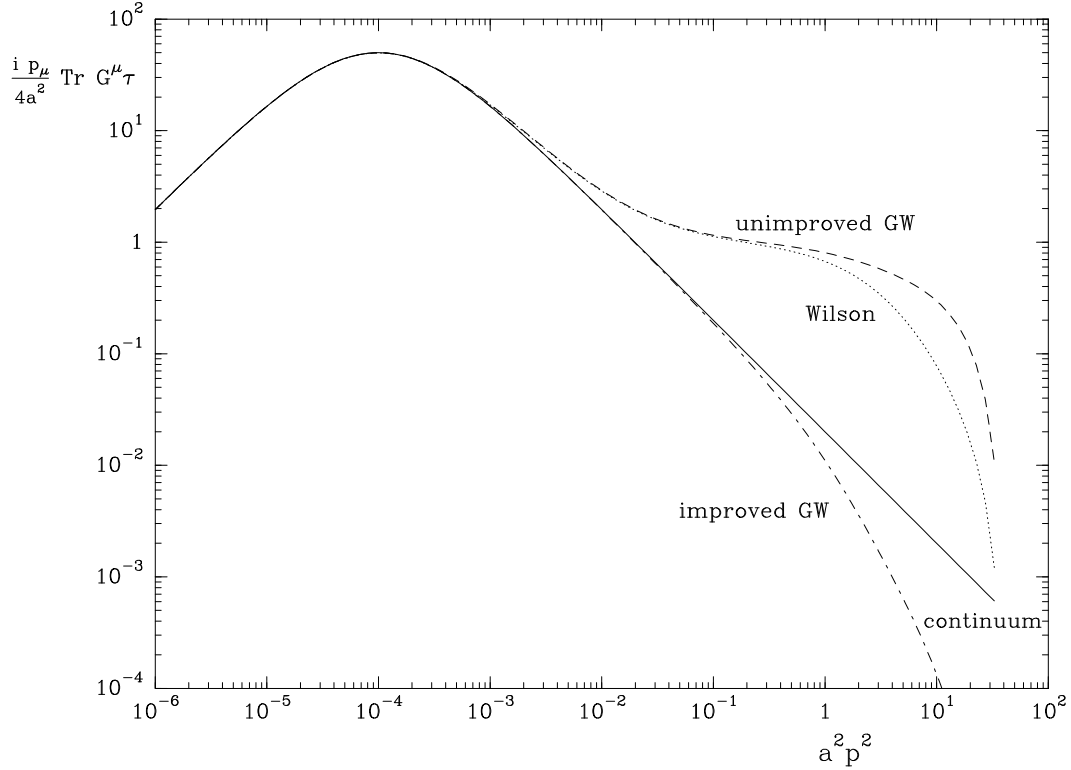, angle = 270, width = 14cm} 
\end{center} 
\vspace*{0.5cm} 
\caption{ 
  The scalar part of the Green's function for the local 
 vector current ${\rm i} p_\mu /(4 a^2){\rm Tr}[ G^\mu(p) \tau]$. 
 The Wilson fermion gives the dotted line, the result 
 from Ginsparg-Wilson fermions with no improvement terms
 is the dashed line, the dot-dashed line is from the improved 
 Green's function $G_\star$, and the solid line is the 
 continuum result  $2 m_0 p^2/(p^2 + m_0^2)^2$. 
 As in Fig.~\ref{fig:prop} all curves are plotted for
 $a m_0 = 0.01$,  and  
 the momentum is taken in the direction $(1,1,1,1)$. 
  }
\label{fig:gammamu}
\end{figure}

 As with the propagator, we can illustrate the effects of improving
 the Green's function by looking at a simple case in the free theory, 
 namely the Green's function $G^\mu(p,m_0)$ for the
 local vector current $\bar{\psi} \gamma_\mu \tau \psi$. Here 
 $\tau$ is a flavour matrix, ${\rm Tr}\, \tau = 0$, normalised
 so that ${\rm Tr}\, \tau^2 = 1$. As in the
 case of the propagator, the most sensitive test is to look 
 at the scalar part of the Green's function, which is shown 
 in Fig.~\ref{fig:gammamu}. We have chosen the quantity 
 \begin{equation}
 \sum_\mu {\rm i} \frac{ \sin a p_\mu}{a}
  \frac{1}{4 a^2} {\rm Tr} \left[ G^\mu(p,m_0) \, \tau \right], 
 \end{equation} 
 which has the value $2 m_0 p^2/(p^2 + m_0^2)^2$
 in the continuum. Again, the Wilson action and the unimproved
 Ginsparg-Wilson Green's function deviate significantly from
 the desired continuum result at $a^2 p^2 \approx a m_0$, 
 while the improved Green's function only has errors of $O(a^2)$
 and remains reliable until $a^2 p^2 \approx 1$.    

\begin{figure}[tb]
\begin{center}
\epsfig{file = 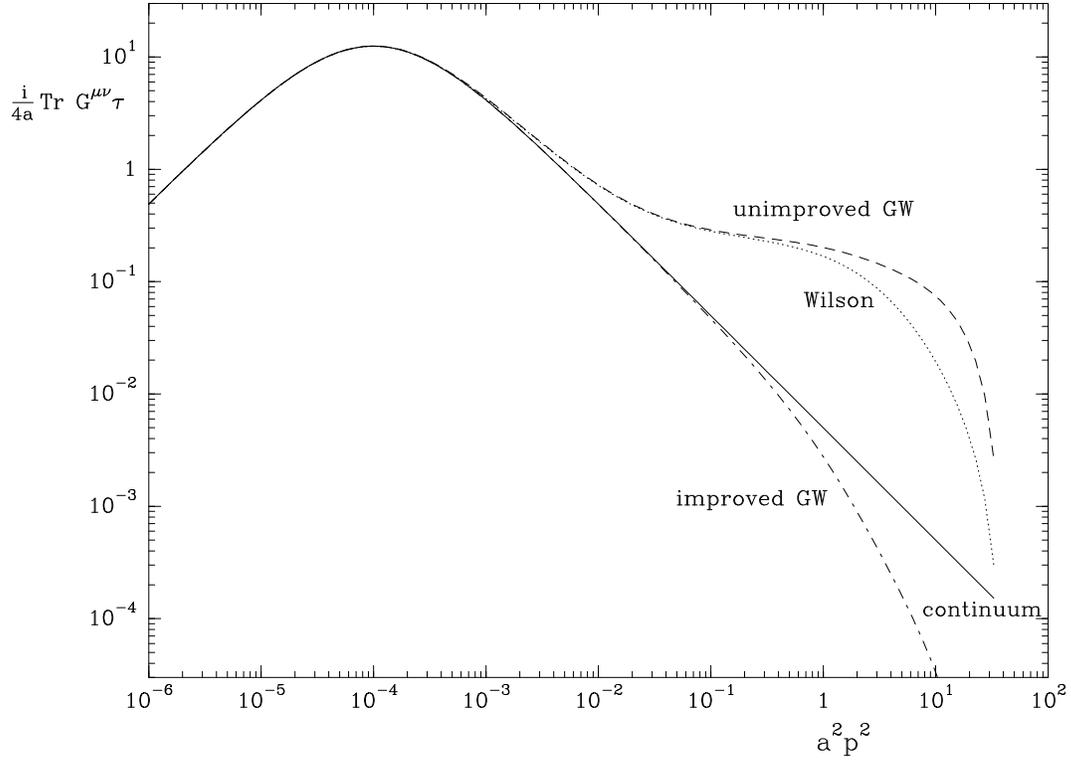, angle = 270, width = 14cm}
\end{center}
\vspace*{0.5cm}
\caption{
 The scalar part of the Green's function for the
 operator $O^{\mu \nu}$ of eq.~(\ref{omunu}),
 ${\rm i} /(4 a) {\rm Tr} [ G^{\mu \nu}(p, m_0) \tau ]$.
 The Wilson fermion gives the dotted line, the result
 from Ginsparg-Wilson fermions with no improvement terms
 is the dashed line, the dot-dashed line is from the improved
 Green's function $G_\star$, and the solid line is the
 continuum result  $2 m_0 p^\mu p^\nu /(p^2 + m_0^2)^2$.
 As in Fig.\ref{fig:prop} all curves are plotted for
 $a m_0 = 0.01$  and
 the momentum is taken in the direction $(1,1,1,1)$.
  }
\label{fig:freex}
\end{figure}

   As a final example in the free case, let us take an operator
 with a derivative, namely
 \begin{equation}
 O^{\mu \nu} \equiv \frac{ \rm i}{2} \bar{\psi} \, \frac{1}{2}
 \left( \gamma_\mu \Dlr_\nu + \gamma_\nu \Dlr_\mu \right) \tau \, \psi , 
 \label{omunu}
 \end{equation}
 with $\mu \ne \nu$. This is an operator that can be used to compute
 the flavour non-singlet moment $\langle x \rangle$ of a hadronic structure
 function. In the continuum, the free Green's function for this operator
 has the value
 \begin{equation}
 G^{\mu \nu}_{\rm cont}(p, m_0)
 = \frac{ -{\rm i} \pslash + m_0 }{p^2 + m_0^2} \,
 \frac{1}{2} \left( \gamma_\mu p_\nu + \gamma_\nu p_\mu \right)
 \, \tau \, \frac{ -{\rm i} \pslash + m_0 }{p^2 + m_0^2} .
 \end{equation}
 The improved lattice Green's function found by applying
 the formulae in sect.~\ref{sect:impop} is
  \begin{equation}
 G^{\mu \nu}_\star(p, m_0) =
 S_\star(p, m_0)  \,
 \frac{1}{2 a}  \left( \gamma_\mu \sin a p_\nu +
  \gamma_\nu \sin a p_\mu \right) \, \tau \,
  S_\star(p, m_0) ,
 \end{equation}
 where the explicit form of $S_\star(p, m_0)$ is given in
 eq.~(\ref{Sstarfree}). Since the discretisation errors of
 $S_\star$ are $O(a^2)$ it is clear that the difference between
 $ G^{\mu \nu}_{\rm cont}(p, m_0) $ and $ G^{\mu \nu}_\star(p, m_0)$
 is also of $O(a^2)$. This is illustrated in Fig.~\ref{fig:freex} ,
 where we plot the trace of $G^{\mu \nu}$.

 \section{Ward Identities} 

 In the continuum chiral symmetry relates the scalar part of the 
 fermion propagator to the forward Green's function for the 
 non-singlet pseudoscalar operator $\bar{\psi} \gamma_5 \tau \psi$: 
 \begin{equation}
 \gamma_5 \tau S(p) + S(p) \gamma_5 \tau = 2 m G^5(p) \, . 
 \label{contWard} 
 \end{equation}  
 On the lattice we can check that 
 the improved Green's functions of sect.~\ref{sect:impop}
 satisfy the same Ward identity: 
 \begin{equation}
 \gamma_5 \tau S_\star(p) + S_\star(p) \gamma_5 \tau 
 = 2 m_0 G^5_\star(p) \, . 
 \label{lattWard} 
 \end{equation}  
 One might naively expect that this Ward identity would be 
 violated at $O(a^2)$ by the discretisation errors, but
 in fact it holds exactly, even though
 both the left hand side and the right hand side
 have discretisation errors of $O(a^2)$. 

  Note that all the Green's functions needed for 
 Ward identities, like eq.~(\ref{lattWard}), can be computed on 
 the lattice. So we can hope to use them to determine the
 improvement coefficients in other improved theories, such as  
 clover fermions, where we cannot calculate them 
 analytically from first principles. 

 \section{Conclusions}

   We see that using the Ginsparg-Wilson action there are many 
 ways of constructing propagators and Green's functions which
 are correct to $O(a^2)$, both on-shell and off-shell. 
 In this paper we have considered both, improvement by 
 adding the Green's functions of irrelevant operators,
 and by rotating the fermion fields. 

  The idea is very simple. In eqs.~(\ref{Sheur}) and (\ref{Ggoal}) 
 we have the Green's functions we wish to obtain, with the
 correct chiral symmetries, and therefore free from all $O(a)$ 
 errors. However, there they are expressed in terms of the 
 associated matrix $K_{GW}$, defined in eq.~(\ref{kdef}). 
 Fortunately, it is easy to find equivalent expressions 
 in terms of the well-defined matrix $M$, so we never need to 
 find or invert $K_{GW}$ explicitly.

   The improvement coefficients are universal. It makes no difference
 which theory is being considered, Abelian or non-Abelian gauge theory, 
 or a theory with no gauge symmetry, or what value of the coupling is
 being used. It does not even matter which operator is being considered. 
 Operator improvement for other improved actions, such as clover
 fermions, will certainly not be so simple. The improvement coefficients
 in that case will certainly be functions of the coupling, and depend
 on the theory considered.  Nevertheless, Ginsparg-Wilson fermions may
 give some hints as to what sort of improvement terms are needed, and
 what sort of contact terms are to be expected.

 \end{document}